\documentclass[prd,twocolumn,nopacs,floatfix,amsmath,nofootinbib,amssymb,floatfix]{revtex4}
\usepackage{graphicx,color,dcolumn,booktabs,bm}
\usepackage{longtable,lscape}
\usepackage{pdfpages}
\usepackage{txfonts}
\usepackage{overpic}
\usepackage{amssymb}
\usepackage{indentfirst}
\usepackage{feynmf}   
\usepackage{slashed}  
\usepackage{cases}
\usepackage{color}
\usepackage{multirow}
\usepackage{threeparttable}
\usepackage{epstopdf}
\usepackage{enumerate}
\usepackage{graphicx,color,dcolumn,booktabs,bm}
\usepackage[colorlinks, citecolor=blue,anchorcolor=red,menucolor=red, linkcolor=red,filecolor=red,urlcolor=blue,frenchlinks=red]{hyperref}

\graphicspath{{Figures/}} %

\begin{document}

\title{Constructing the $J^{P(C)}=1^{-(+)}$ light flavor hybrid nonet with the newly observed $\eta_1(1855)$}

\author{Bing Chen$^{1,3}$}\email{chenbing@ahstu.edu.cn}
\author{Si-Qiang Luo$^{2,3,4,6}$}\email{luosq15@lzu.edu.cn}
\author{Xiang Liu$^{2,3,4,5,6}$}\email{xiangliu@lzu.edu.cn}
\affiliation{ $^1$School of Electrical and Electronic Engineering, Anhui Science and Technology University, Bengbu 233000, China\\
$^2$School of Physical Science and Technology, Lanzhou University, Lanzhou 730000, China\\
$^3$Lanzhou Center for Theoretical Physics, Key Laboratory of Theoretical Physics of Gansu Province, Lanzhou University, Lanzhou, Gansu 730000, China\\
$^4$Key Laboratory of Quantum Theory and Applications of MoE, Lanzhou University,
Lanzhou 730000, China\\
$^5$MoE Frontiers Science Center for Rare Isotopes, Lanzhou University, Lanzhou 730000, China\\
$^6$Research Center for Hadron and CSR Physics, Lanzhou University and Institute of Modern Physics of CAS, Lanzhou 730000, China}

\date{\today}

\begin{abstract}
The recently discovered $\eta_1(1855)$ and the previously observed $\pi_1(1600)$ state present a valuable opportunity for the investigation of the $J^{P(C)}=1^{-(+)}$ light hybrid nonet. In this study, we employ a semirelativistic quark potential model to examine the masses of the $J^{P(C)}=1^{-(+)}$ light hybrid states. The static potential, which portrays the confinement force between the quark-antiquark pair in a hybrid system, is borrowed from the SU$(3)$ lattice gauge theory.
Additionally, we utilize a constituent gluon model to analyze the strong decay characteristics of these light $1^{-+}$ hybrids. Our findings suggest that the $\pi_1(1600)$ and $\eta_1(1855)$ states could be potential candidates for $1^{-+}$ $(u\bar{u}-d\bar{d})g/\sqrt{2}$ and $s\bar{s}g$ hybrids, respectively. To ensure comprehensiveness, we also investigate the isospin partners of the $\pi_1(1600)$ and $\eta_1(1855)$ states within the $1^{-(+)}$ nonet$-$specifically, the $(u\bar{u}+d\bar{d})g/\sqrt{2}$ and $s\bar{q}g$ ($q=u$ and $d$ quarks) states. We propose some potential decay channels which could be explored in experimental settings to detect these undiscovered states.
\end{abstract}



\maketitle

\section{Introduction}\label{sec1}
Nearly half a century has elapsed since the establishment of quantum chromodynamics (QCD)~\cite{tHooft:1972tcz,Fritzsch:1973pi,Gross:1973id,Politzer:1973fx}. However, physicists are still confronted with the challenge of describing the fundamental properties of hadrons directly from first principles. This difficulty arises primarily from the lack of a clear understanding of the role played by the gluonic field in the low-energy regime of strong interactions (for a comprehensive review of QCD, refer to Ref. \cite{Gross:2022hyw}). Consequently, various phenomenological models, incorporating key aspects of QCD, have been employed to gain insight into the properties of meson and baryon states \cite{Olsen:2017bmm}.

Within the framework of quark potential models, the gluonic field in conventional hadron systems, responsible for mediating the interaction between valence quarks, is typically approximated by an effective adiabatic potential~\cite{Quigg:1979vr,Lucha:1991vn,Mukherjee:1993hb}. By employing this effective potential, the mass spectra of meson and baryon states can be successfully reproduced. However, it is important to note that the quark potential model has not been rigorously examined in the context of hadrons containing an excited gluonic field. These particular types of mesons, which possess an excited gluonic field, are referred to as the hybrid states and are considered to be exotic states.
Undoubtedly, conducting a comprehensive study of hybrid states could significantly contribute to our understanding of the role played by the gluonic field in low-energy strong interactions. Unfortunately, as of now, none of the hybrid states have been definitively established, which underscores the need for further investigation in this area.

According to the results of lattice QCD~\cite{Dudek:2011bn}, the quantum numbers $J^{PC}$ of the lightest hybrid mesons have been proposed to be $0^{-+}$, $1^{-+}$, $2^{-+}$, and $1^{--}$. However, it should be noted that identifying the $0^{-+}$, $2^{-+}$, and $1^{--}$ hybrid states in experiments can be challenging, as they may mix with the conventional $q\bar{q}$ mesons in the nearby mass region. On the other hand, the $1^{-+}$ states have the potential to serve as a distinctive ``smoking gun" signature for the presence of spin-exotic states, provided that their existence is confirmed experimentally. To date, the Particle Data Group (PDG) has cataloged three $1^{-+}$ states, specifically the $\pi_1(1400)$, $\pi_1(1600)$, and $\pi_1(2015)$~\cite{ParticleDataGroup:2022pth}. Among these states, the $\pi_1(2015)$ has solely been observed by the E852 experiment in the decay modes $f_1(1285)\pi$~\cite{E852:2004gpn} and $b_1(1235)\pi$~\cite{E852:2004rfa}. Nevertheless, further substantial experimental evidence is required to establish its existence definitively.


The existence of the $\pi_1(1400)$ state remains a subject of controversy. In various processes such as $\pi^-$ diffraction~\cite{IHEP-Brussels-LosAlamos-AnnecyLAPP:1988iqi,Aoyagi:1993kn,E852:1997gvf,E852:1999xev,VES:2001rwn,Dzierba:2003fw,E862:2006cfp} and $p\bar{p}/n\bar{p}$ annihilation~\cite{CrystalBarrel:1998cfz,CrystalBarrel:1999reg,CrystalBarrel:2019zqh}, the $\pi_1(1400)$ state has been observed with decay modes including $\eta\pi$ and $\rho\pi$. However, in the cascade process $\psi(3686)\to\gamma\chi_{c1}\to\gamma\pi_1(1400)^\pm\pi^\mp\to\gamma\eta\pi^\pm\pi^\mp$ involving approximately $4.48\times10^8$ $\psi(3686)$ events, the BESIII Collaboration did not detect a clear signal for the $\pi_1(1400)$~\cite{BESIII:2016tqo}.
Similarly, the COMPASS Collaboration conducted a comprehensive resonance-model analysis of the $\pi^-\pi^-\pi^+$ invariant mass spectrum in the reaction $\pi^-+p\to\pi^-\pi^-\pi^+ + p_{\textup{spectator}}$ and found no distinct resonance signal for the $\pi_1(1400)$ state\cite{COMPASS:2018uzl}. Moreover, the JPAC Collaboration performed a coupled-channel analysis of COMPASS data for diffractively produced $\eta^{(\prime)}\pi^-$ final states~\cite{JPAC:2018zyd}, revealing only one clear resonance pole corresponding to the $\pi_1(1600)$ state in the spin-exotic wave. Consequently, it is plausible to suggest that the structure attributed to the $\pi_1(1400)$ might be an artifact resulting from imperfections in the analysis methodology. This conclusion finds support in the work of Kopf \emph{et al}.~\cite{Kopf:2020yoa}, where a sophisticated coupled-channel analysis of data from the COMPASS and Crystal Barrel experiments demonstrated that a single $\pi_1(1600)$ pole effectively describes the amplitude of the $1^-(1^{-+})$ wave.


Among the spin-exotic states, the $\pi_1(1600)$ stands as the sole experimental finding thus far. It has been detected in various decay modes, including $\eta^\prime\pi$~\cite{VES:1993scg,E852:2001ikk,Amelin:2005ry}, $\rho\pi$~\cite{E852:1998mbq,Zaitsev:2000rc,COMPASS:2009xrl,COMPASS:2021ogp}, $f_1(1285)\pi$~\cite{E852:2004gpn}, and $b_1(1235)\pi$~\cite{E852:2004rfa,Baker:2003jh}. However, precise measurements of the Breit-Wigner parameters and important branching fractions for the $\pi_1(1600)$ state have been challenging due to substantial background contributions from nonresonant processes. Detailed information regarding the $\pi_1(1400)$, $\pi_1(1600)$, and $\pi_1(2015)$ states can be found in the comprehensive reviews~\cite{Meyer:2010ku,Meyer:2015eta,Ketzer:2019wmd,Chen:2022asf,Gross:2022hyw}. The available data on the $\pi_1(1600)$ state provide support for its candidacy as a hybrid meson. Assuming the $\pi_1(1600)$ to be an isovector hybrid within the $J^{PC}=1^{-+}$ nonet, one can anticipate the potential experimental discovery of its isospin partners.

A significant advancement was recently achieved by the BESIII Collaboration, as they discovered the first isoscalar $1^{-+}$ state, named the $\eta_1(1855)$, through a partial wave analysis of $J/\psi\to\gamma\eta_1(1855)\to\gamma\eta\eta^\prime$~\cite{BESIII:2022iwi,BESIII:2022riz}. The observation of the $\eta_1(1855)$ state has spurred various theoretical interpretations. It has been proposed as a molecular state of $K\bar{K}_1(1400)$ or a dynamically generated state in Refs.~\cite{Dong:2022cuw,Yang:2022rck,Yan:2023vbh}. In the framework of QCD sum rules, the $\eta_1(1855)$ has been suggested as a tetraquark state with the configuration $[1_c]{s\bar{s}}\otimes[1_c]_{q\bar{q}}$~\cite{Wan:2022xkx}. Furthermore, investigations into the $\eta_1(1855)$ as a hybrid candidate have been conducted using the flux tube model~\cite{Qiu:2022ktc}, QCD sum rules~\cite{Chen:2022qpd}, and the effective Lagrangian method~\cite{Shastry:2022mhk}. In addition, the production cross sections of the $\eta_1(1855)$ in reactions such as $K^-p\to\eta(1855)\Lambda$~\cite{Wang:2022sib} and $\gamma p\to\eta_1(1855)p$ reactions~\cite{Huang:2022tpq},  as well as the branching fractions of processes like $J/\psi\to\gamma\eta_1(1855)$~\cite{Shastry:2023ths,Chen:2022isv} and $J/\psi\to\eta^{(\prime)}\eta_1(1855)$~\cite{Yu:2022wtu}, have been extensively studied.


As emphasized in Ref.~\cite{Gross:2022hyw}, establishing the complete SU(3)$_{\textup{flavor}}$ multiplet for the $1^{-+}$ states is an essential endeavor. The observations of the $\pi_1(1600)$ and $\eta_1(1855)$ states provide an excellent opportunity to pursue this research. In this study, we investigate the masses of $1^{-+}$ hybrid states in Sec. \ref{sec2}, utilizing a hybrid static potential simulated by lattice gauge theory. Furthermore, we analyze the strong decays of the $\eta_1(1855)$, $\pi_1(1600)$, and their isospin partners within the $1^{-+}$ nonet. The decay model is introduced in Sec. \ref{sec3}, and the obtained results are presented in Sec. \ref{sec4}. Finally, we end the paper with discussion and conclusion in Sec. \ref{sec5}.


\section{Masses of $1^{-+}$ hybrid nonet}\label{sec2}

\begin{figure}[htbp]
\centering
\includegraphics[width=8.4cm,keepaspectratio]{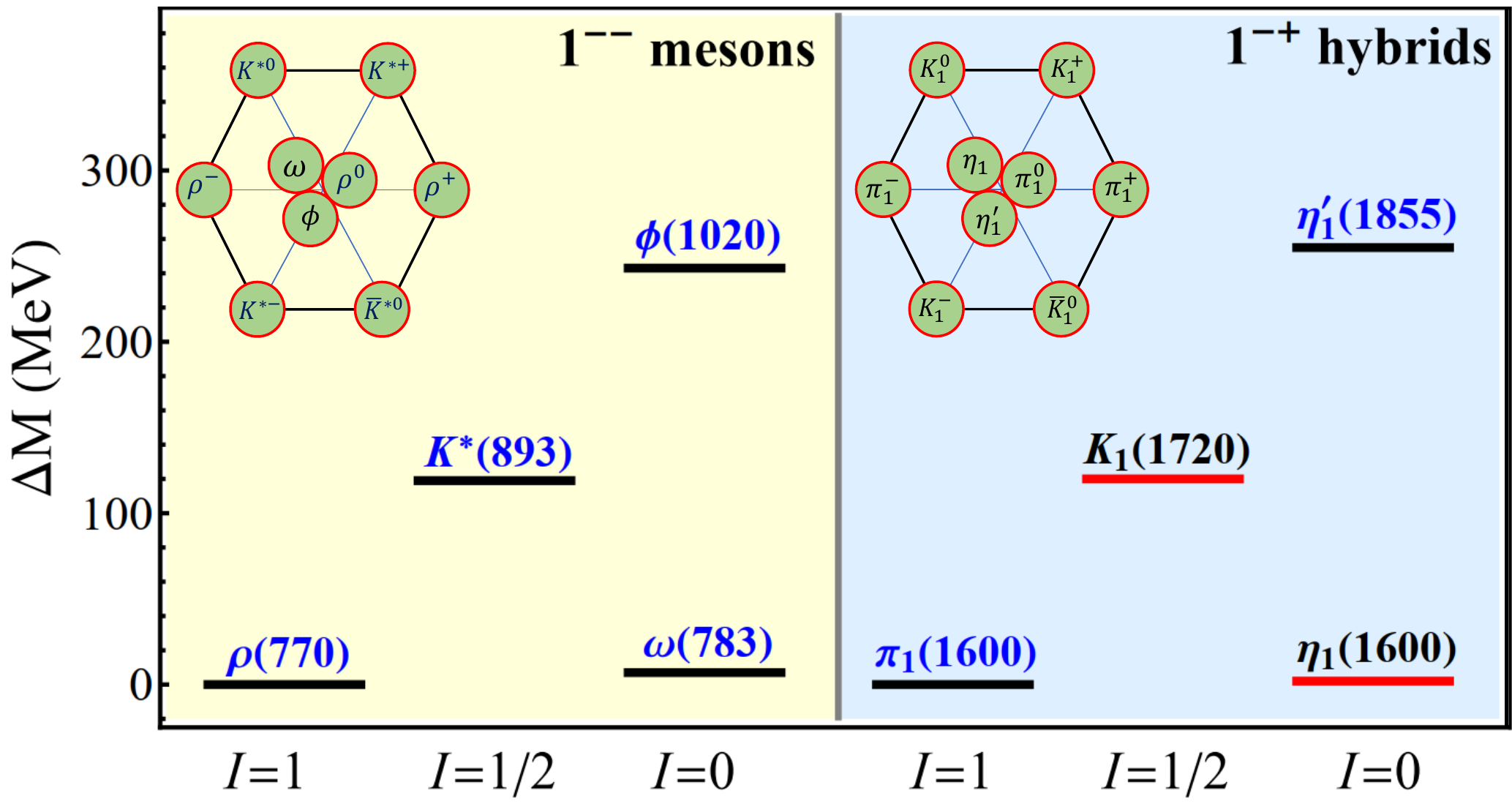}
\caption{Mass differences of $\phi(1020)/\rho(770)$ and $\eta_1^\prime(1855)/\pi_1(1600)$. For the light mesons, $\Delta{m}$ is defined as $m_X-m_\rho$ where $X$ denotes a state in the $1^3S_1$ nonet. For the light hybrids, $\Delta{m}$ is defined as $m_X-m_{\pi_1}$ where $X$ denotes a state in the $1^{-+}$ nonet. The $\omega(783)$, $K^\ast(893)$, and the unknown $\eta_1(1600)/K_1(1720)$ are also presented. The masses of the $\rho(770)$, $\omega(783)$, $K^\ast(893)$, and $\phi(1020)$ states are taken from Ref.~\cite{ParticleDataGroup:2022pth}. The masses of the $\pi_1(1600)$ and $\eta_1^\prime(1855)$ states are taken from Refs.~\cite{COMPASS:2018uzl,BESIII:2022iwi,BESIII:2022riz}.}
\label{massgaps}
\end{figure}

It is worth noting that the mass difference between the $\eta_1(1855)$ and $\pi_1(1600)$ states is remarkably similar to that between the $\phi(1020)$ and $\rho(770)$ mesons (as shown in Fig. \ref{massgaps}). This observation can be explained as follows: If we consider the $\pi_1(1600)$ and $\eta_1(1855)$ as $1^{-+}$ hybrids, the spin of the constituent quark-antiquark pair, denoted as $S_{q\bar{q}}$, is equal to 1, which is the same as for the $\rho(770)$ and $\phi(1020)$ mesons. The primary distinction lies in the fact that the $q\bar{q}$ pair in the $\rho(770)$ and $\phi(1020)$ systems is associated with the lowest-lying static potential, while the $q\bar{q}$ pair in the $\pi_1(1600)$ and $\eta_1(1855)$ systems is formed from the lowest excited configurations of the gluon field. Under this assumption, the mass gaps observed between $\phi(1020)/\rho(770)$ and $\eta_1(1855)/\pi_1(1600)$ can be predominantly attributed to the inherent mass difference between the $s$ and $u/d$ quarks.

To further investigate the hybrid nature of the $\pi_1(1600)$ and $\eta_1^\prime(1855)$ states,\footnote{Since the $1^{-+}$ isoscalar state observed by BESIII~\cite{BESIII:2022iwi,BESIII:2022riz} is proposed to be the strange isoscalar partner of the $\pi_1(1600)$ state, we henceforth refer to it as the $\eta_1^\prime(1855)$ state.} we employ a quark potential model to calculate the masses of these lowest hybrid states. The quark potential model, which has been validated by lattice QCD simulations~\cite{Barkai:1984ca,Itoh:1985wk}, has exhibited considerable success in describing the mass spectrum of meson and baryon systems. Consequently, it can be extended to incorporate hybrid masses. Notably, the interaction between quarks, mediated by the excited gluon field, has been simulated as ``effective potentials" in lattice QCD studies~\cite{UKQCD:1998zbe,Karbstein:2018mzo,Capitani:2018rox,Schlosser:2021wnr}.


In the subsequent analysis, we utilize a quark potential model to compute the masses of $1^{-+}$ light hybrid states, incorporating the hybrid static potential extracted from lattice simulations~\cite{Capitani:2018rox}.\footnote{Reference~\cite{Capitani:2018rox} provides more precise hybrid static potentials with smaller statistical errors and finer spatial resolution for different gluon field configurations.} It is worth noting that lattice QCD not only obtains effective potentials for hybrids but also successfully reproduces the well-established Cornell potential for conventional meson states. For the $q\bar{q}$ mesons, the effective potential is
\begin{equation}
V_{\Sigma_g^+}(r)=-\frac{4}{3}\frac{\alpha_s}{r}+br+V_0.  \label{eq1}
\end{equation}
The parameters $\alpha$, $b$, and $C$ stand for the strength of the color Coulomb potential, the strength of linear confinement, and a mass-renormalized constant, respectively. For the lowest-energy hybrids, the effective potential is given as
\begin{equation}
V_{\Pi_u}(r)=\frac{A_1}{r}+A_2r^2+V_1.   \label{eq2}
\end{equation}
On the other hand, the short-distance behavior of the gluonic field $\Pi_u$ can be accurately described by a $1^{+-}$ gluelump, as discussed in the Ref. \cite{Braaten:2014qka}. This understanding has received further confirmation through lattice QCD computations, which have shown that the gluonic field in the lowest-energy hybrids exhibits a significant overlap with the chromomagnetic structure characterized by $J^{PC}_g=1^{+-}$ \cite{Dudek:2011bn}.
When considering a $q\bar{q}$ pair in an internal $S$-wave, the permitted $J^{PC}$ values for the lightest hybrid mesons are $0^{-+}$, $1^{-+}$, $2^{-+}$, and $1^{--}$. These conclusions align with the findings of lattice QCD calculations \cite{Dudek:2011bn}, where the masses of the $0^{-+}$, $1^{-+}$, $2^{-+}$, and $1^{--}$ hybrid mesons were expected to be of the order of
\begin{equation}
M_{0^{-+}}<M_{1^{-+}}<M_{1^{--}}<M_{2^{-+}}.\label{eq3}
\end{equation}

Here, we are only concerned about the masses of $1^{-+}$ hybrids. In our calculations, the following spinless Salpeter equation\footnote{The spinless Salpeter equation could be regarded as an approximate version of the relativistic Bethe-Salpeter equation~\cite{Salpeter:1951sz,Salpeter:1952ib}. More details can be found in Ref.~\cite{Lucha:1998ix}.} is solved for the masses of the light mesons and the lowest hybrid states:
\begin{equation}
\left[\sqrt{m_1^2+p^2}+\sqrt{m_2^2+p^2}+V(r)\right]\psi_{nL}=E\psi_{nL}. \label{eq4}
\end{equation}
For mesons, the potential $V(r)$ in Eq.~(\ref{eq2}) includes not only the $V_{\Sigma_g^+}(r)$ in Eq.~(\ref{eq1}), but also the spin-spin contact hyperfine interaction,
\begin{equation}
V_{q_1\bar{q}_2}^{\textup{cont.}}(r)=\frac{32\alpha\sigma^3e^{-\sigma^2r^2}}{9\sqrt{\pi}m_1m_2}\textbf{s}_{q_1}\cdot\textbf{s}_{\bar{q}_2}. \label{eq5}
\end{equation}
The parameters in Eqs.(\ref{eq1}) and (\ref{eq5}) can be determined by the measured masses of well-established $q\bar{q}$ states. Then the parameters of the potential $V_{\Pi_u}(r)$ for the lowest hybrid states can be constrained by the following relationships: $A_1=0.0958$, $A_2=\xi_2(b/\xi_1)^{3/2}$, and $V_1=V_0+\xi_3\sqrt{b/\xi_1}$. Here, we have $\xi_1=0.04749$, $\xi_2=0.001599$, and $\xi_3=0.5385$.
In principle, the potential of hybrid mesons should include the spin-spin interaction. However, due to the current limitations in experimental data, the scarcity of measurements restricts our ability to thoroughly discuss the spin-spin effects of these hybrids. As a result, our primary focus in this work centers aon determining the central masses of $0^{-+}$, $1^{-+}$, $2^{-+}$, and $1^{--}$ hybrids.

\begin{table}[htbp]
\caption{Predicted masses of $n^{2S+1}L_J$ light mesons with the corresponding states (GeV).
}\label{table1}
\renewcommand\arraystretch{1.15}
\begin{tabular*}{86mm}{c@{\extracolsep{\fill}}cccccc}
\toprule[1pt]\toprule[1pt]
\multirow{2}{*}{$n^{2S+1}L_{(J)}$}   & \multicolumn{2}{c}{$q\bar{q}$ meson}  & \multicolumn{2}{c}{$q\bar{s}$ meson}  & \multicolumn{2}{c}{$s\bar{s}$ meson}    \\
\cline{2-3}\cline{4-5}\cline{6-7}
                  &  Pred$.$     &   Exp$.$~\cite{ParticleDataGroup:2022pth}       &   Pred$.$  &   Exp$.$~\cite{ParticleDataGroup:2022pth}       & Pred$.$  &   Exp$.$~\cite{ParticleDataGroup:2022pth}  \\
\toprule[1pt]
  $1^1S_0$        &   0.139      &   $\pi(140)$   &    0.490   & $K(496)$       & 0.754   & $\eta^\prime(958)$ \\
  $1^3S_1$        &   0.772      &   $\rho(775)$  &    0.897   & $K^\ast(892)$  & 1.018   & $\phi(1020)$   \\
  $2^1S_0$        &   1.107      &  $\pi(1300)$   &    1.342   & $K(1460)$      & 1.540   & $\eta^\prime(1475)$ \\
  $2^3S_1$        &   1.397      &  $\rho(1450)$  &    1.529   & $K^\ast(1410)$ & 1.659   & $\phi(1680)$  \\
  $1^1P_1$        &   1.072      &  $h_1(1170)$   &    1.286   & $K_1(1270)$    & 1.465   & $h_1(1415)$  \\
  $1^3P_J$        &   1.262      &  $a_1(1260)$   &    1.413   & $K_1(1400)$    & 1.553   & $f_2^\prime(1525)$  \\
  $1^1D_2$        &   1.559      & $\eta_2(1645)$ &    1.722   & $K_2(1770)$    & 1.871   &               \\
  $1^3D_J$        &   1.611      & $\rho_3(1690)$ &    1.761   & $K_3(1780)$    & 1.900   & $\phi_3(1850)$ \\
\bottomrule[1pt]\bottomrule[1pt]
\end{tabular*}
\end{table}

By reproducing the masses of these well-determined $q\bar{q}$ mesons, as shown in Table \ref{table1}, we fix the parameters of the potential model as follows: The masses of the $u/d$ and $s$ quarks are taken to be 0.32 GeV and 0.45 GeV, respectively. The parameters $\alpha_s$ and $b$ are set to 0.64 and 0.165 GeV$^2$ for all light mesons. The values for $\sigma$ and $V_0$ are assigned as follows: $\sigma_{q\bar{q}}=0.47$ GeV, $\sigma_{q\bar{s}}=0.45$ GeV, $\sigma_{s\bar{s}}=0.43$ GeV, $V_0^{(q\bar{q})}=-0.48$ GeV, $V_0^{(q\bar{s})}=-0.40$ GeV, and $V_0^{(s\bar{s})}=-0.33$ GeV. With these parameter values, the average mass of the $\Pi_u$ hybrid multiplet is summarized in the bottom row of Table \ref{table2}, which also includes the newly measured masses of the $\pi_1(1600)$ and $\eta_1^\prime(1855)$ states for comparison.


\begin{table}[htbp]
\caption{A comparison of the predicted average masses of the $\Pi_u$ hybrid multiplet with the measured masses of the $\pi_1(1600)$ and $\eta_1^\prime(1855)$ states (MeV).} \label{table2}
\renewcommand\arraystretch{1.2}
\begin{tabular*}{86mm}{@{\extracolsep{\fill}}ccc}
\toprule[1pt]\toprule[1pt]
$\pi_1(1600)$      &  $K_1$     &  $\eta_1^\prime(1855)$  \\
\toprule[1pt]
  $1600^{+110}_{-60}$~\cite{COMPASS:2018uzl}    &    &  $1855\pm9^{+6}_{-1}$~\cite{BESIII:2022riz}  \\
  $1564\pm24\pm86$~\cite{JPAC:2018zyd}          &    &           \\
  $1623\pm47^{+24}_{-75}$~\cite{Kopf:2020yoa}   &    &       \\
\toprule[1pt]
   1669       &        1852     & 2023                 \\
\bottomrule[1pt]\bottomrule[1pt]
\end{tabular*}
\end{table}

As presented in Table \ref{table2}, the measured mass of the $\pi_1(1600)$ state is comparable to the predicted average mass of the lowest $q\bar{q}g$ hybrids, suggesting that the $\pi_1(1600)$ could indeed be a viable $1^{-+}$ hybrid state. In fact, our prediction for the lowest $q\bar{q}g$ hybrids aligns well with most theoretical predictions for the $1^{-+}$ $q\bar{q}g$ state (see Fig. \textcolor{red}{78} of Ref. \cite{Chen:2022asf} for a comparison). Furthermore, several lattice QCD calculations also support the hybrid assignment for the $\pi_1(1600)$ state, particularly when extrapolating the pion mass in the lattice QCD simulations to the physical value \cite{Chen:2022asf}. Thus, the mass of the $\pi_1(1600)$ state provides a compelling evidence in favor of its hybrid nature.

Regarding the predicted average mass for the $s\bar{s}g$ states, it is consistent with the findings of Refs. \cite{Isgur:1984bm,Close:2003ae,Eshraim:2020ucw}, but approximately 150 MeV higher than the measured mass of the $\eta_1^\prime(1855)$ state. However, it is important to note that our calculations do not account for spin-dependent interactions. Thus, the hybrid assignment for the $\eta_1^\prime(1855)$ remains a plausible possibility. Swanson recently studied the flavor mixing of hybrid states and also proposed the $\eta_1^\prime(1855)$ state as the probable $s\bar{s}g$ partner of the $\pi_1(1600)$ state \cite{Swanson:2023zlm}.

Certainly, further investigation is warranted to explore the possibility of the $\pi_1(1600)$ and $\eta_1^\prime(1855)$ states as two hybrid members of the SU(3)$_{\textup{flavor}}$ $1^{-+}$ multiplet. In the following analysis, we will examine their strong decay behavior under the assumption of the $1^{-+}$ hybrid assignment. To calculate the strong decays of the $\pi_1(1600)$ and $\eta_1^\prime(1855)$ states, we will employ a constituent gluon model proposed in Ref.~\cite{Farina:2020slb}. For the spatial wave functions of the initial hybrid and final meson states, we will approximate them using the simple harmonic oscillator (SHO) wave function. By adopting this approach, we can obtain analytical expressions for the decay amplitudes. In our calculations, we will determine the scale parameter $\beta$ of the SHO wave function by solving the Salpeter equation. The specific values of $\beta$ for the final meson states are collected in Table \ref{table3}. Additionally, we present the $\beta^H_{q_1\bar{q}_2}$ values (where $q_1$ and $q_2$ represent the $u$, $d$, and $s$ quarks) for the lowest hybrid states in the $\Pi_u$ multiplet as follows:
\begin{equation}
\begin{aligned}
\beta^H_{q\bar{q}}=0.264 ~\textup{GeV},~~~\beta^H_{q\bar{s}}=0.271 ~\textup{GeV},~~~\beta^H_{s\bar{s}}=0.277 ~\textup{GeV}. \nonumber
\end{aligned}
\end{equation}
The predicted values of $\beta^H_{q\bar{q}}$ and $\beta^H_{q\bar{s}}$ are consistent with the findings of the flux-tube model~\cite{Close:2003ae}, indicating agreement between the two approaches. However, the predicted value of $\beta^H_{s\bar{s}}$ is slightly smaller than the corresponding result from the flux-tube model.

\begin{table}[htbp]
\caption{Values of the SHO wave function scale $\beta$ (GeV).
}\label{table3}
\renewcommand\arraystretch{1.16}
\begin{tabular*}{84mm}{c@{\extracolsep{\fill}}cccccc}
\toprule[1pt]\toprule[1pt]
  System          &  $1^1S_0$  &  $1^3S_1$  &   $2^1S_0$  &  $2^3S_1$  &    $1^1P_1$  &  $1^3P_J$       \\
\toprule[1pt]
  $q\bar{q}$      &  0.669     & 0.396      & 0.435       &   0.356    & 0.446        &   0.339 \\
  $q\bar{s}$      &  0.609     & 0.432      & 0.421       &   0.371    & 0.421        &   0.353 \\
  $s\bar{s}$      &  0.579     & 0.466      & 0.417       &   0.384    & 0.412        &   0.367 \\
\bottomrule[1pt]\bottomrule[1pt]
\end{tabular*}
\end{table}


\section{Decay model of the hybrid state}\label{sec3}

Numerous models and methods have been proposed to investigate the strong decays of hybrid states, including the constituent gluon model~\cite{Tanimoto:1982eh,Close:2003ae,Tanimoto:1982wy,LeYaouanc:1984gh,Iddir:1988jd,Ishida:1991mx,Kalashnikova:1993xb,Swanson:1997wy,Iddir:2000yb,Ding:2006ya,Iddir:2007dq,Benhamida:2019nfx,Farina:2020slb}, the flux-tube model~\cite{Isgur:1985vy,Close:1994hc,Barnes:1995hc,Page:1998gz}, QCD sum rules~\cite{DeViron:1984svx,Zhu:1998sv,Zhang:2002id,Chen:2010ic,Huang:2010dc,Huang:2016upt}, lattice QCD~\cite{McNeile:2002az,McNeile:2006bz,Woss:2020ayi}, and other approaches~\cite{Page:2003qn,Shastry:2022mhk}. In this study, we employ the constituent gluon model, recently developed in Ref. \cite{Farina:2020slb}, to calculate the strong decays of $1^{-+}$ light hybrid states. This model has successfully reproduced reasonable spin-averaged $c\bar{c}g$ spectra, which are in good agreement with lattice QCD results \cite{HadronSpectrum:2012gic}. In the following, we will demonstrate that the constituent gluon model can also capture the main decay characteristics of the $\pi_1(1600)$ state predicted by lattice QCD~\cite{Woss:2020ayi}.

In the constituent gluon model, the gluon dynamics in the lowest hybrid states is approximated by an axial gluon with quantum numbers $J^{PC}=1^{+-}$. The foundation of this model can be understood as follows: in these lowest hybrid states, the close proximity of the quark-antiquark pair allows the gluon field configuration to be effectively described by the corresponding gluelumps \cite{Braaten:2014qka}. Consequently, a $1^{-+}$ hybrid state can be simplified as a three-body system, with the quark-antiquark pair and a transverse electric (TE) gluon $(J_g^{PC}=1^{+-})$ as its essential degrees of freedom.

\begin{figure}[htbp]
\centering
\includegraphics[width=4.2cm,keepaspectratio]{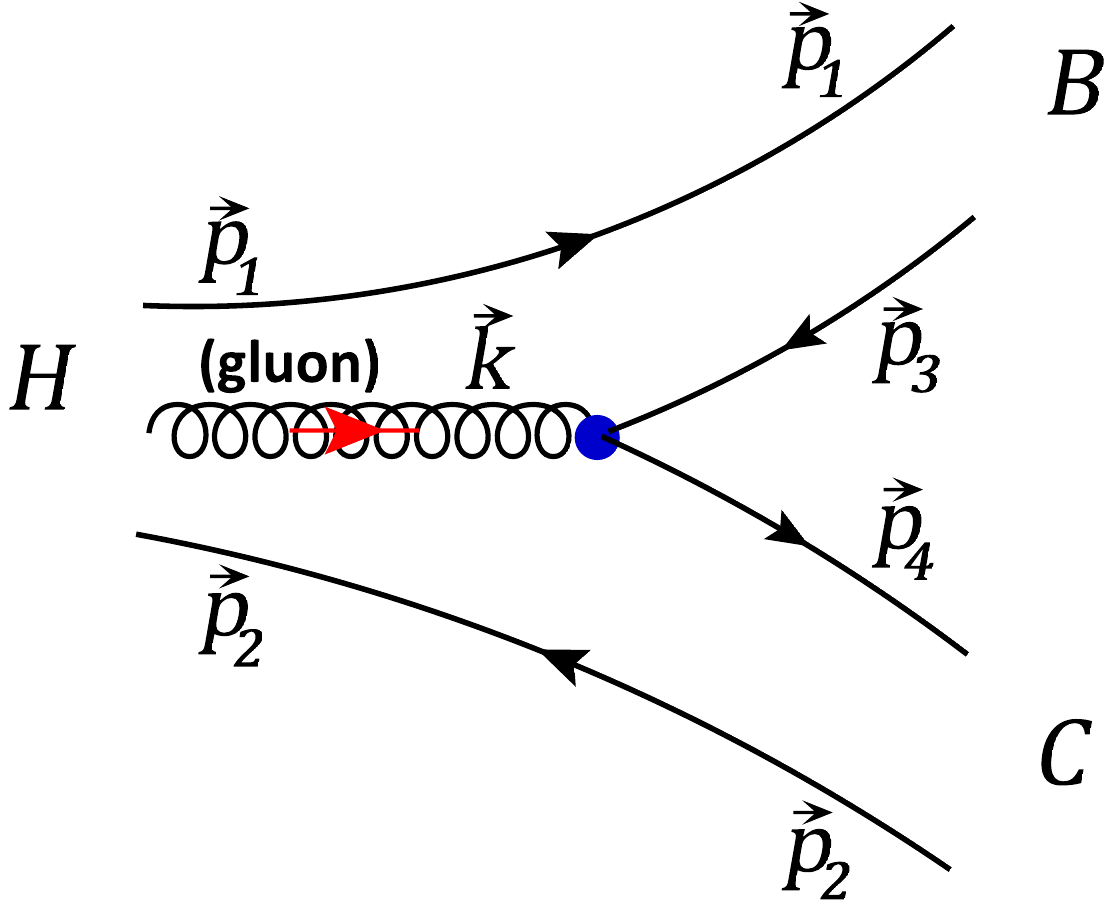}
\caption{A connected diagram for a hybrid state $H$ decaying into the $B$ and $C$ mesons.}
\label{decay}
\end{figure}


A matrix element of the interaction Hamiltonian in the QCD theory, which describes the annihilation of a gluon and the creation of a quark-antiquark pair, is taken as the coupling vertex to describe a decay process of a hybrid state into two normal mesons. As shown in Fig. \ref{decay}, the corresponding Hamiltionian is given by
\begin{equation}
\hat{H}_I=g_s\int\textup{d}^3\vec{x}~\bar{\psi}(\vec{x})\gamma_\mu\frac{\lambda^a}{2}\psi(\vec{x})A_a^\mu(\vec{x})  \label{eq6}
\end{equation}
in the lowest order~\cite{LeYaouanc:1984gh}. The transition operator $\hat{\mathcal{T}}$ could be derived from the above matrix element as
\begin{equation}
\begin{split}
\hat{\mathcal{T}}=&g_s\sum_{s,s^\prime,\lambda}\int\frac{\textup{d}^3\vec{p}_3\textup{d}^3\vec{p}_4\textup{d}^3\vec{k}}{\sqrt{2\omega_g}(2\pi)^6}\delta^3(\vec{p}_3+\vec{p}_4-\vec{k})\\
&\times\frac{\lambda_{c_g}^{c,c^\prime}}{2}\omega_g^{(34)}\chi_s^\dagger\sigma\tilde{\chi}_{s^\prime}\epsilon^\mu(\hat{\textbf{k}},\lambda)b_3^\dagger(\vec{p}_3)d_4^\dagger(\vec{p}_4)a_{\textbf{k}\lambda}^{c_g}(\vec{k})  \label{eq7}
\end{split}
\end{equation}
in the nonrelativistic limit. The parameter $\omega_g$ could be regarded as the effective mass of the constituent gluon~\cite{Swanson:1997wy}, which is taken to be 0.80 GeV in our calculations. The strong coupling constant $g_s$ is constrained by the decay width of $\pi_1(1600)$ with the $1^{-+}$ hybrid assignment.
In practical application, the mock state is adopted to describe the spatial wave function of initial and final hadrons~\cite{Hayne:1981zy}. For the hybrid state, the wave function could be written as
\begin{equation}
\begin{split}
|\textup{H}(J_Hm_H~[&L_H(L_{q_1\bar{q}_2}J_g)S_H](\vec{P}_H)\rangle\equiv\\
&\omega_H\phi_H\mathcal{W}_{\textup{CG}}\int \textup{d}^3\vec{p}_1\textup{d}^3\vec{p}_2\textup{d}^3\vec{k}\delta^3(\vec{p}_1+\vec{p}_2+\vec{k}-\vec{P}_H)\\
&\times\Psi_{n_{q_1\bar{q}_2},n_g}^{L_{q\bar{q}}m_{q\bar{q}},L_gm_g}(\vec{p}_1,\vec{p}_2,\vec{k})|q_1(\vec{p}_1)\bar{q}_2(\vec{p}_2)g(\vec{k})\rangle. \label{eq8}
\end{split}
\end{equation}
where $\omega_H$ and $\phi_H$ denote the color and flavor wave functions, respectively, of an initial hybrid state. As a color singlet state, the color wave functions of a hybrid state could be represented as $\omega_H=|(q\bar{q})_8\otimes{g_8}\rangle^0$. The flavor wave function of a hybrid state is the same as that of a meson state, since the gluon is a flavor singlet. The value $\mathcal{W}_{\textup{CG}}$ in Eq.~(\ref{eq4}) involves the Clebsch-Gordan coefficients, which denote the spins and angular momentums of the quarks and gluon coupling to the total spin ($J_H$) of a hybrid state. Specifically, it is given as
\begin{equation}
\begin{split}
\mathcal{W}_{\textup{CG}}=&\sqrt{\frac{2J_g+1}{4\pi}}\mathcal{D}^{J_g\ast}_{m_g,\mu}\chi_{\mu,\lambda}\langle{L_{q_1\bar{q}_2}m_{q_1\bar{q}_2},J_gm_g}|L_Hm_L\rangle\\
&\times\langle\frac{1}{2}m_1,\frac{1}{2}m_2|S_Hs_H\rangle\langle{S_Hs_H,L_Hm_L}|J_Hm_H\rangle. \label{eq9}
\end{split}
\end{equation}
The value $\chi_{\mu,\lambda}$ denotes the spin of a transverse gluon in the gluon helicity basis$-$i.e. $\chi_{\mu,\lambda}\equiv\langle{1\lambda,L_g0}|J_g\mu\rangle\delta_{\mu\lambda}$. For the $1^{-+}$ hybrid states containing a TE gluon, the factor $\chi_{\mu,\lambda}$ is always $1/\sqrt{2}$. The role of the rotation matrix $\mathcal{D}^{J_g\ast}_{m_g,\mu}$ in Eq.~(\ref{eq5}) is to convert the angular momentum projection of a gluon to the basis of the hybrid system. As done in Refs.~\cite{LeYaouanc:1984gh,Iddir:1988jd,Farina:2020slb}, a simple product ansatz could be made for the spatial wave function of the hybrid state. Namely, the $\Psi_{n_{q\bar{q}}}^{L_{q\bar{q}},L_g}(\vec{p}_1,\vec{p}_2,\vec{k})$ could be written as
\begin{equation}
\Psi_{n_{q\bar{q}},n_g}^{L_{q\bar{q}},L_g}(\vec{p}_1,\vec{p}_2,\vec{k})=\psi_{n_{q_1\bar{q}_2}}^{L_{q\bar{q}}m_{q\bar{q}}}(\vec{p}_\rho)\psi_{n_g}^{L_gm_g}(\vec{p}_\lambda). \label{eq10}
\end{equation}
With the Jacobian coordinates, $\vec{p}_\rho$ and $\vec{p}_\lambda$ are represented as follows:
\begin{equation}
\vec{p}_\rho=\frac{1}{2}\vec{p}-\vec{p}_B-\frac{1}{2}\frac{m_1-m_2}{m_1+m_2}\vec{k};~~~~\vec{p}_\lambda=\vec{k} \label{eq11}
\end{equation}
in the rest frame of the hybrid system. The values $m_1$ and $m_2$ refer to the masses of $q_1$ and $\bar{q}_2$ in the hybrid state. As mentioned before, the SHO wave function could be used for the spatial wave function $\psi_n^{Lm}(\vec{p})$. When the wave functions of the final meson states in the decay process are constructed in the same way, the momentum $\vec{p}$ and $\vec{k}$ can be integrated by performing the partial wave amplitude calculation. In this way, the helicity amplitude of a decay process $H\to{B+C}$ (see Fig. \ref{decay}) can be obtained analytically by the relation $\langle{BC}|\hat{\mathcal{T}}|H\rangle=\delta^3(\textbf{K}_B+\textbf{K}_C)\mathcal{M}^{j_H,j_B,j_C}(\textbf{p})$. The concrete expression of $\mathcal{M}^{j_H,j_B,j_C}(\textbf{p})$ can be found in Ref. \cite{Farina:2020slb}.


Finally, the partial wave amplitude is given as
\begin{equation}
\begin{aligned}\label{eq12}
\mathcal {M}^{H\to BC}_{LS}=\frac{\sqrt{2L+1}}{2J_H+1}&\sum_{\text{$j_B$,$j_C$}}\langle L0Jj_H|J_Hj_A\rangle\\
&\times\langle J_Bj_B,J_Cj_C|Jj_H\rangle\mathcal
{M}^{j_H,j_B,j_C}(\textbf{p}),
\end{aligned}
\end{equation}
while the partial width of the process $H\to BC$ in the $H$ rest frame is given by
\begin{equation}
\begin{aligned}\label{eq12}
\Gamma(H\rightarrow BC)=2\pi\frac{E_BE_C}{M_H}p\sum_{L,S}|\mathcal
{M}^{H\to BC}_{LS}(p)|^2.
\end{aligned}
\end{equation}


\section{Strong decays of the $1^{-+}$ nonet}\label{sec4}

\subsection{The $\pi_1(1^{-+})$ state}

\begin{table}[htbp]
\caption{The partial and total widths of the strong decays of the $\pi_1(1^{-+})$ state (MeV). Here, the final mesons refer to all allowed charge conjugate pairs. For example, when we consider $\pi_1(1600)^-$ decaying into the $K^\ast(892)$ and $K$ states, $K^\ast{K}$ denotes the $K^\ast(892)^-+K^0$ and $K^\ast(892)^0+K^-$ channels. If a partial width is predicted to be smaller than 1 MeV, we denote it as zero.} \label{table4}
\renewcommand\arraystretch{1.2}
\begin{tabular*}{86mm}{@{\extracolsep{\fill}}ccccc}
\toprule[1pt]\toprule[1pt]
 $\rho\pi$       & $K^\ast{K}$     & $b_1(1235)\pi$   & $f_1(1285)\pi$    &   $f_2(1270)\pi$  \\
 2               & $\approx 0$     &  244             & 15                &   $\approx 0$            \\
\cline{4-5}
                 &                 &                  & Total             & Exp.~\cite{ParticleDataGroup:2022pth}   \\
                 &                 &                  & 261               & 240$\pm$50  \\
\bottomrule[1pt]\bottomrule[1pt]
\end{tabular*}
\end{table}

By assigning $\pi_1(1600)$ as the isovector member in the $J^{PC}=1^{-+}$ nonet, we give a total decay width of 261 MeV,\footnote{By comparing with the average decay width of $\pi_1(1600)$ listed in the PDG table, we may find that the uncertainty of our result is about $10\%\sim30\%$. The same uncertainty also exists in the results of Tables \ref{table5}, \ref{table6}, and \ref{table7}.} which is comparable to the average value reported in the PDG table~\cite{ParticleDataGroup:2022pth}.\footnote{It should be noted that the pole position (or the Breit-Wigner parameter) of the $\pi_1(1600)$ state currently has a large uncertainty, and more precise measurements are required in the future.} The dominant decay modes $b_1(1260)\pi$ and $f_1(1280)\pi$ are predicted. This finding is consistent with the results of the flux-tube model~\cite{Close:1994hc} and lattice QCD~\cite{Woss:2020ayi}, where the $b_1(1260)\pi$ channel was also identified to have the largest partial decay width for the $\pi_1(1600)$ state. The decays of $\pi_1(1600)$ into the $b_1(1260)\pi$ and $f_1(1280)\pi$ channels can occur via both $s$-wave and $d$-wave processes. However, our calculations show that the $d$-wave partial widths are quite negligible ($<$ 1 MeV) due to the limited phase space available. This conclusion is in agreement with the lattice QCD results~\cite{Woss:2020ayi}, but it contradicts the predictions of the flux-tube model~\cite{Close:1994hc}.

In addition to the $b_1(1260)\pi$ and $f_1(1280)\pi$ channels, the $\pi_1(1600)$ state has also been observed in the $\rho(770)\pi$ \cite{E852:1998mbq,Zaitsev:2000rc,COMPASS:2009xrl,COMPASS:2021ogp} and $\eta^\prime(958)\pi$ \cite{VES:1993scg,E852:2001ikk,Amelin:2005ry} channels. According to the results presented in Table \ref{table4}, the partial decay width for the $\rho\pi$ channel is estimated to be approximately 2 MeV. This suggests that the $\pi_1(1600)$ state could be observed in the $\pi_1(1600)\to\rho\pi$ decay process, given a sufficiently large data sample. However, the isospin symmetry \cite{Close:1987aw} prohibits the decays of $\pi_1(1600)$ into $\eta\pi$ and $\eta^\prime\pi$ via the mechanism shown in Fig. \ref{decay}. Therefore, this conclusion contradicts the observation of $\pi_1(1600)\to\eta^\prime\pi$ reported in previous studies~\cite{VES:1993scg,E852:2001ikk,Amelin:2005ry}. One possible explanation for this puzzle is that the $\eta^\prime$ meson may have a significant gluonium component\footnote{The KLOE Collaboration determined the gluonium content of the $\eta^\prime$ meson through a global fit to the radiative decays of pseudoscalar and vector mesons~\cite{Ambrosino:2009sc}. Their analysis indicated a substantial gluonium component in the wave function of the $\eta^\prime$ meson.}. Consequently, the decay mechanism of $\pi_1(1600)\to\eta^\prime\pi$ would differ from the other decay modes listed in Table \ref{table4}. Further discussion on this point will be provided in the next subsection.

We would like to emphasize the significance of the branching ratio $\mathcal{B}=\frac{\Gamma(\pi_1(1600)\to b_1(1235)\pi)}{\Gamma(\pi_1(1600)\to f_1(1285)\pi)}$ in providing insights into the nature of the $\pi_1(1600)$ state. When considering the mass and SHO wave function scale parameter $\beta$ of the $b_1(1235)$ and $f_1(1285)$ to be equal, the ratio $\mathcal{B}$ is determined to be 4, a result consistent with the findings of the flux-tube model \cite{Burns:2006wz}. However, by incorporating the differences between the $b_1(1235)$ and $f_1(1285)$ in the calculation, the value of $\mathcal{B}$ increases to approximately 16.3. This suggests that if the $\pi_1(1600)$ is indeed a $1^{-+}$ hybrid state, its partial width for $\pi_1(1600)\to b_1(1235)\pi$ is expected to be significantly larger than that for $\pi_1(1600)\to f_1(1285)\pi$. Currently, only the VES experiment has evaluated the ratio $\mathcal{B}$ with a large uncertainty \cite{Amelin:2005ry}. Therefore, further experimental efforts are needed to measure this branching ratio accurately.


\subsection{The $\eta_1(1^{-+})$ state}

As the isoscalar counterpart to the $\pi_1(1600)$ state, it is anticipated that the $\eta_1(1^{-+})$ state will have a mass approximately equal to that of the $\pi_1(1600)$ state \cite{Eshraim:2020ucw}. However, the decay width of the $\eta_1(1^{-+})$ state is predicted to be narrow (see Table \ref{table5}), which is in stark contrast to the broader decay width of the $\pi_1(1600)$. The same result has also been presented in Ref. \cite{Shastry:2023ths} where the lighter isoscalar hybrid was predicted to be the narrowest state in the $1^{-+}$ multiplet.

Table \ref{table5} reveals that the dominant decay channel for the $\eta_1(1^{-+})$ state is $a_1(1260)\pi$. Therefore, a promising avenue for detecting the $\eta_1(1^{-+})$ signal is through the $J/\psi\to\gamma \eta_1(1^{-+})\to\gamma+a_1(1260)\pi$ process, as $J/\psi$ radiative decays are known to be rich in hybrid production. However, the situation is different for the $\pi p$ diffraction process at high beam energy, as this process is primarily governed by Pomeron exchange \cite{Ketzer:2019wmd}. Consequently, the production of $\eta_1(1^{-+})$ in $\pi p$ reactions is suppressed.

\begin{table}[htbp]
\caption{The partial and total widths of the strong decays of the $\eta_1(1^{-+})$ state (MeV).} \label{table5}
\renewcommand\arraystretch{1.2}
\begin{tabular*}{86mm}{@{\extracolsep{\fill}}ccccc}
\toprule[1pt]\toprule[1pt]
 $\eta\eta^\prime$ & $K^\ast{K}$    &  $a_1(1260)\pi$  & $a_2(1320)\pi$     & $\pi(1300)\pi$      \\
 $\approx 0$       & $\approx 0$    &   55             & $\approx 0$        & 5                \\
\cline{4-5}
                   &                &                  &  Total             &  Exp.    \\
                   &                &                  & 60                 &  $-$      \\
\bottomrule[1pt]\bottomrule[1pt]
\end{tabular*}
\end{table}

In the scenario where the $\eta_1(1^{-+})$ hybrid contains a small $s\bar{s}g$ component,\footnote{The mixing between two isoscalar states in $1^{-+}$ nonet is expected to be small due to the homochiral nature \cite{Shastry:2023ths}.} a visible mass difference between the $\eta_1(1^{-+})$ and $\pi_1(1600)$ states could be expected. However, considering the decay properties of a pure $1^{-+}$ $s\bar{s}g$ state discussed in the next subsection, it is evident that the $a_1(1260)\pi$ channel remains the dominant decay mode for the $\eta_1(1^{-+})$ state when its mass is below 1.76 GeV. Consequently, the main partial widths predicted and listed in Table \ref{table5} would not undergo significant changes.


\subsection{The $\eta^\prime_1(1^{-+})$ state}

As discussed earlier, the mass difference between the $\eta_1^\prime(1855)$ and $\pi_1(1600)$ states can be naturally explained by treating the $\eta_1^\prime(1855)$ state as the $s\bar{s}g$ partner of the $\pi_1(1600)$. In this section, we will initially consider the $\eta_1^\prime(1855)$ as a pure $s\bar{s}g$ hybrid state and examine its strong decays. Subsequently, we will discuss the possibility of the $\eta_1^\prime(1855)$ state being a mixture of $s\bar{s}g$ and $(u\bar{u}+d\bar{d})g/\sqrt{2}$ components.

Treating the $\eta_1^\prime(1855)$ state as a $s\bar{s}g$ hybrid, the total decay width is predicted to be approximately 160 MeV, which agrees with experimental measurements~\cite{BESIII:2022iwi,BESIII:2022riz}. However, the partial decay width of $\eta\eta^\prime$ is estimated to be no more than 1 MeV, which may appear too small to be detected in experiments. This result is analogous to the case of $\pi_1(1600)\to\eta^\prime\pi$. The presence of gluonium content in the $\eta^\prime$ meson can significantly influence the $\eta^\prime(1855)\to\eta\eta^\prime$ process. In other words, the decay mechanism governing the $\eta^\prime(1855)\to\eta\eta^\prime$ channel differs from that of other channels listed in Table \ref{table6}. If the $\pi_1(1600)$ and $\eta_1^\prime(1855)$ are $1^{-+}$ hybrids, there should be a nearby $0^{-+}$ isoscalar hybrid state. This $0^{-+}$ isoscalar hybrid state can mix with the $\eta^\prime$ meson, resulting in the gluonic content in the $\eta^\prime$. Experimental measurements of the branching ratio $\mathcal{B}(\eta_1^\prime(1855)\to\eta\eta^\prime)/\mathcal{B}(\eta_1^\prime(1855)\to K^\ast K)$ can test this speculation in the future. Taking into account the gluonic content of the $\eta^\prime$, the partial width of $\eta_1^\prime(1855)\to\eta\eta^\prime$ is comparable to or even larger than that of $\eta_1^\prime(1855)\to K^\ast K$.

\begin{table}[htbp]
\caption{The partial and total widths of the strong decays of the $\eta_1^\prime(1^{-+})$ state (MeV).} \label{table6}
\renewcommand\arraystretch{1.2}
\begin{tabular*}{86mm}{@{\extracolsep{\fill}}ccccc}
\toprule[1pt]\toprule[1pt]
 $\eta\eta^\prime$          &   $K^\ast{K}$       & $K_1(1270)K$     & Total      &  Exp.~\cite{BESIII:2022iwi,BESIII:2022riz}  \\
  $\approx 0$               &    2                &  157             & 159        &  188$\pm$18$^{+3}_{-8}$  \\
\bottomrule[1pt]\bottomrule[1pt]
\end{tabular*}
\end{table}

It is also crucial for future experiments to search for the $\eta_1^\prime(1855)$ state in the $\eta_1^\prime(1855)\to K_1(1270)K$ process since $K_1(1270)K$ is predicted to be a dominant decay channel for the $\eta_1^\prime(1855)$ state. For the decay channels associated with $K_1(1270)$ and $K_1(1400)$, it is necessary to regard them as a combination of the $K_1(^1P_1)$ and $K_1(^3P_1)$ states. The mixing scheme reads as \cite{Suzuki:1993yc}
\begin{equation}
\left\{
\begin{array}{c}
|K_1(1270)\rangle = \cos\theta~|K_1(^1P_1)\rangle - \sin\theta~|K_1(^3P_1)\rangle\\
~|K_1(1400)\rangle = \sin\theta~|K_1(^1P_1)\rangle + \cos\theta~|K_1(^3P_1)\rangle.
\end{array}
\right.
\end{equation}
The mixing angle $\theta$ is adopted as $60^\circ$, a value established through an examination of the $K_1\to K\pi\pi$ strong decays \cite{Tayduganov:2011ui}.

As an isoscalar meson, $\eta_1^\prime(1855)$ in principle could contain some $(u\bar{u}+d\bar{d})g/\sqrt{2}$ component \cite{Swanson:2023zlm}. In future, experiments can check the $(u\bar{u}+d\bar{d})g/\sqrt{2}$ component of the $\eta_1^\prime(1855)$ state by searching for it in the $a_1(1260)\pi$ channel.


\subsection{The $K_1(1^{-})$ state}

Given the predicted decay width of 278 MeV, the $K_1(1^-)$ state, as the strange partner of the $\pi_1(1600)$ and $\eta^\prime(1855)$ states, is anticipated to be broad. Consequently, it may pose a challenge for future experimental searches, as detecting such a broad state can be more challenging compared to the narrower resonances. Nonetheless, further investigations and experimental efforts are crucial to exploring the properties and existence of the $K_1(1^-)$ state.

\begin{table}[htbp]
\caption{The partial and total widths of the strong decays of the $K_1(1^-)$ state where the mass of the $K_1(1^-)$ state is taken as 1.72 GeV (MeV).} \label{table7}
\renewcommand\arraystretch{1.2}
\begin{tabular*}{86mm}{@{\extracolsep{\fill}}ccccc}
\toprule[1pt]\toprule[1pt]
 $K\pi$           & $K\eta$          & $K\eta^\prime$  & $K\rho$         & $K\omega$   \\
 1                & $\approx 0$      & $\approx 0$     & $\approx 0$     & $\approx 0$    \\
 $K^\ast\pi$      & $K^\ast\eta$     & $K^\ast\rho$    & $K^\ast\omega$  & $K_1(1270)\pi$   \\
  3               & 1                & $\approx 0$     & $\approx 0$     & 106          \\
 $K_1(1400)\pi$   & $K_2(1430)\pi$   & $h_1(1116)K$    &$K(1460)\pi$     & $K^\ast(1410)\pi$  \\
 146              & $\approx 0$      & 16              & 2               & 3       \\
\cline{4-5}
                  &                  &                 &  Total          &  Exp.   \\
                  &                  &                 &  278            &    $-$     \\
\bottomrule[1pt]\bottomrule[1pt]
\end{tabular*}
\end{table}

The $K_1(1^-)$ state is predicted to exhibit significant decays into $K_1(1400)\pi$ and $K_1(1270)\pi$, making them the two dominant decay channels (see Table \ref{table7}). Furthermore, the decays $h_1(1170)K$, $K^\ast(1410)\pi$, and $K^\ast\pi$ are also expected to be observable for the $K_1(1^-)$ state. Given that the $\pi_1(1600)$ state has been observed in the $\eta^\prime\pi$ channel and the $\eta_1^\prime(1855)$ state in the $\eta^\prime\eta$ channel, we propose that future experimental investigations also target the $K_1(1^-)$ state in the $K\eta^\prime$ channel. These suggested explorations would provide valuable insights into the existence and properties of the $K_1(1^-)$ state.

\section{Discussion and conclusion}\label{sec5}

In the field of hadron physics, the study of exotic states has gained significant attention. While flavor-exotic states like the $Z_c^\pm$ and $Z_b^\pm$ have been extensively investigated, spin-exotic states hold a special significance in understanding the nonperturbative nature of strong interactions at low energies. Currently, there have been reports of three $1^{-+}$ states: the $\pi_1(1400)$, $\pi_1(1600)$, and $\pi_1(2015)$. However, the existence of the $\pi_1(1400)$ and $\pi_1(2015)$ states is still subject to debate, while the existence of the $\pi_1(1600)$ is relatively more certain. Even so, crucial properties such as the pole position and important decay ratios of the $\pi_1(1600)$ state are yet to be precisely measured. Consequently, the establishment of the light $J^{P(C)}=1^{-(+)}$ hybrid nonet is still a work in progress.

Recently, the discovery of the $\eta^\prime_1(1855)$ state by the BESIII Collaboration in the $\eta\eta^\prime$ decay channel has significantly impacted the situation~\cite{BESIII:2022iwi,BESIII:2022riz}. As both the $\eta^\prime(1855)$ and $\pi_1(1600)$ are $1^{-+}$ states, their mass difference closely matches that of the $\phi$ and $\rho$ mesons. This observation suggests that the $\eta^\prime(1855)$ and $\pi_1(1600)$ could be two members of a $J^{P(C)}=1^{-(+)}$ hybrid nonet. By assuming the $\eta^\prime(1855)$ to be the $s\bar{s}g$ partner of the $\pi_1(1600)$, we conducted a study on their masses and strong decays. The results indicate that the $\pi_1(1600)$ and $\eta^\prime(1855)$ are plausible candidates to be considered as the members of $1^{-+}$ hybrid nonet.

The predicted dominant decay modes of the $\pi_1(1600)$ state include $b_1(1260)\pi$ and $f_1(1280)\pi$, which is consistent with experimental observations since the $\pi_1(1600)$ has been detected in these two decay channels. To further determine the nature of the $\pi_1(1600)$ state, future experiments could focus on measuring the ratio of $\Gamma(\pi_1(1600)\to b_1(1235)\pi)$ to $\Gamma(\pi_1(1600)\to f_1(1285)\pi)$.
The predicted partial width for the $\rho\pi$ channel is around 2 MeV, in agreement with lattice QCD results~\cite{Woss:2020ayi}. However, the $\eta^\prime\pi$ channel is forbidden for the $\pi_1(1600)$ if the $\eta^\prime$ is considered a conventional $q\bar{q}$ meson. This discrepancy seems to contradict experimental observations where the $\pi_1(1600)$ state has been found in the $\eta^\prime\pi$ channel~\cite{VES:1993scg,E852:2001ikk,Amelin:2005ry}. We have proposed that the decay mechanism of $\pi_1(1600)\to\eta^\prime\pi$ might be different from other observed decay processes due to the significant gluonic component of the $\eta^\prime$ meson. This assumption also provides an explanation for the observation of $\eta^\prime_1(1855)\to\eta^\prime\eta$. In fact, decay processes involving the $\eta^\prime$ meson in the final states are considered as promising channels to search for hybrid states \cite{Bibrzycki:2021rwh}. Therefore, it is crucial to measure the partial widths of $\pi_1(1600)\to\eta^\prime\pi$ and $\eta^\prime_1(1855)\to\eta^\prime\eta$ more precisely in future experiments.


Being isospin partners of the $\pi_1(1600)$ and $\eta_1^\prime(1855)$ in the $1^{-+}$ hybrid multiplet, the $K_1$ state is expected to have a broad decay width, which presents a challenge for experimental searches. In contrast, the $\eta_1$ state could potentially be observed in the cascade process $J/\psi\to\gamma \eta_1(1^{-+})\to\gamma+a_1(1260)\pi$. In the future, experiments such as BESIII \cite{BESIII:2020nme}, Belle II \cite{Belle-II:2018jsg}, GlueX \cite{Dudek:2012vr,Burkert:2018nvj}, PANDA \cite{PANDA:2009yku}, and COMPASS \cite{Ketzer:2019wmd} will provide more valuable data to enhance our understanding of hybrid states. Therefore, the field of hybrids warrants increased attention from both experimentalists and theorists.

\section*{Acknowledgements}

This project is supported by the National Natural Science Foundation of China under Grants No. 11305003, No. 12047501, and No. 12247101. X. L. is also supported
by China National Funds for Distinguished Young Scientists under Grant No. 11825503, the National Key Research, the Development Program of China under Contract No. 2020YFA0406400, the Fundamental Research Funds for the Central Universities, and the Project for Top-notch Innovative Talents of Gansu Province.



\begin{thebibliography}{99}

\bibitem{tHooft:1972tcz}
  G.~'t Hooft and M.~J.~G.~Veltman,
  Regularization and renormalization of gauge fields,
  Nucl. Phys. B \textbf{44}, 189 (1972).

\bibitem{Fritzsch:1973pi}
  H.~Fritzsch, M.~Gell-Mann and H.~Leutwyler,
  Advantages of the color octet gluon picture,
  Phys. Lett. B \textbf{47}, 365 (1973).

\bibitem{Gross:1973id}
  D.~J.~Gross and F.~Wilczek,
  Ultraviolet behavior of nonabelian gauge theories,
  Phys. Rev. Lett. \textbf{30}, 1343 (1973).

\bibitem{Politzer:1973fx}
  H.~D.~Politzer,
  Reliable perturbative results for strong interactions?,
  Phys. Rev. Lett. \textbf{30}, 1346 (1973).

\bibitem{Gross:2022hyw}
  F.~Gross, E.~Klempt, S.~J.~Brodsky, \textit{et al.}
  50 Years of Quantum Chromodynamics,
  arXiv:2212.11107.

\bibitem{Olsen:2017bmm}
  S.~L.~Olsen, T.~Skwarnicki and D.~Zieminska,
  Nonstandard heavy mesons and baryons: Experimental evidence,
  Rev. Mod. Phys. \textbf{90}, 015003 (2018).

\bibitem{Quigg:1979vr}
  C.~Quigg and J.~L.~Rosner,
  Quantum mechanics with applications to quarkonium,
  Phys. Rep. \textbf{56}, 167 (1979).

\bibitem{Lucha:1991vn}
  W.~Lucha, F.~F.~Schoberl and D.~Gromes,
  Bound states of quarks,
  Phys. Rept. \textbf{200}, 127 (1991).

\bibitem{Mukherjee:1993hb}
  S.~N.~Mukherjee, R.~Nag, S.~Sanyal, T.~Morii, J.~Morishita and M.~Tsuge,
  Quark potential approach to baryons and mesons,
  Phys. Rept. \textbf{231}, 201 (1993).

\bibitem{Dudek:2011bn}
  J.~J.~Dudek,
  The lightest hybrid meson supermultiplet in QCD,
  Phys. Rev. D \textbf{84}, 074023 (2011).


\bibitem{ParticleDataGroup:2022pth}
  R.~L.~Workman \textit{et al.} (Particle Data Group),
  Review of Particle Physics,
  Prog. Theor. Exp. Phys. \textbf{2022}, 083C01 (2022).

\bibitem{E852:2004gpn}
  J.~Kuhn \textit{et al.} (E852 Collaboration),
  Exotic meson production in the $f_1(1285)\pi^-$ system observed in the reaction $\pi^-p\to\eta\pi^+\pi^-\pi^-p$ at 18 GeV$/c$,
  Phys. Lett. B \textbf{595}, 109 (2004).

\bibitem{E852:2004rfa}
  M.~Lu \textit{et al.} (E852 Collaboration),
  Exotic meson decay to $\omega\pi^0\pi^-$,
  Phys. Rev. Lett. \textbf{94}, 032002 (2005)

\bibitem{IHEP-Brussels-LosAlamos-AnnecyLAPP:1988iqi}
  D.~Alde \textit{et al.},
  Evidence for a $1^{-+}$ Exotic Meson,
  Phys. Lett. B \textbf{205}, 397 (1988).

\bibitem{Aoyagi:1993kn}
  H.~Aoyagi \textit{et al.},
  Study of the eta pi- system in the $\pi^-p$ reaction at 6.3 GeV$/c$,
  Phys. Lett. B \textbf{314}, 246 (1993).

\bibitem{E852:1997gvf}
  D.~R.~Thompson \textit{et al.} (E852 Collaboration),
  Evidence for exotic meson production in the reaction $\pi^-p\to\eta\pi^-p$ at 18 GeV$/c$,
  Phys. Rev. Lett. \textbf{79}, 1630 (1997).

\bibitem{E852:1999xev}
  S.~U.~Chung \textit{et al.} (E852 Collaboration),
  Evidence for exotic $J^{PC}=1^{-+}$ meson production in the reaction $\pi^-p\to\eta\pi^-p$ at 18 GeV$/c$,
  Phys. Rev. D \textbf{60}, 092001 (1999).

\bibitem{VES:2001rwn}
  V.~Dorofeev \textit{et al.} (VES Collaboration),
  The $J^{PC}=1^{-+}$ hunting season at VES,
  AIP Conf. Proc. \textbf{619}, 143 (2002)

\bibitem{Dzierba:2003fw}
  A.~R.~Dzierba,\textit{et al.}
  A Study of the $\eta\pi^0$ spectrum and search for a $J^{PC}=1^{-+}$ exotic meson,
  Phys. Rev. D \textbf{67}, 094015 (2003).

\bibitem{E862:2006cfp}
  G.~S.~Adams \textit{et al.} (E852 Collaboration),
  Confirmation of the $1^{-+}$ exotics in the $\eta\pi^0$ system,
  Phys. Lett. B \textbf{657}, 27 (2007).

\bibitem{CrystalBarrel:1998cfz}
  A.~Abele \textit{et al.} (Crystal Barrel Collaboration),
  Exotic $\eta\pi$ state in $\bar{p}d$ annihilation at rest into $\pi^-\pi^0\eta p_{\textup{spectator}}$,
  Phys. Lett. B \textbf{423}, 175 (1998).

\bibitem{CrystalBarrel:1999reg}
  A.~Abele \textit{et al.} (Crystal Barrel Collaboration),
  Evidence for a $\pi\eta$ $P$-wave in $p\bar{p}$ annihilations at rest into $\pi^0\pi^0\eta$,
  Phys. Lett. B \textbf{446}, 349 (1999).

\bibitem{CrystalBarrel:2019zqh}
  M.~Albrecht \textit{et al.} (Crystal Barrel Collaboration),
  Coupled channel analysis of ${\bar{p}p}\,\rightarrow \,\pi ^0\pi ^0\eta $, ${\pi ^0\eta \eta }$ and ${K^+K^-\pi ^0}$ at 900 MeV/c and of ${\pi\pi}$-scattering data,
  Eur. Phys. J. C \textbf{80}, 453 (2020).

\bibitem{BESIII:2016tqo}
  M.~Ablikim \textit{et al.} (BESIII Collaboration),
  Amplitude analysis of the $\chi_{c1} \to \eta\pi^+\pi^-$ decays,
  Phys. Rev. D \textbf{95}, 032002 (2017).

\bibitem{COMPASS:2018uzl}
  M.~Aghasyan \textit{et al.} (COMPASS Collaboration),
  Light isovector resonances in $\pi^- p \to \pi^-\pi^-\pi^+ p$ at 190 GeV/${\it c}$,
  Phys. Rev. D \textbf{98}, 092003 (2018).

\bibitem{JPAC:2018zyd}
  A.~Rodas \textit{et al.} (JPAC Collaboration),
  Determination of the pole position of the lightest hybrid meson candidate,
  Phys. Rev. Lett. \textbf{122}, 042002 (2019).

\bibitem{Kopf:2020yoa}
  B.~Kopf \textit{et al.},
  Investigation of the lightest hybrid meson candidate with a coupled-channel analysis of ${{\bar{p}}p}$-, $\pi ^- p$- and ${\pi \pi }$-Data,
  Eur. Phys. J. C \textbf{81}, 1056 (2021).

\bibitem{VES:1993scg}
  G.~M.~Beladidze \textit{et al.} (VES Collaboration),
  Study of $\pi^-N\to\eta\pi^-N$ and $\pi^-N\to\eta^\prime\pi^-N$ reactions at 37 GeV$/c$,
  Phys. Lett. B \textbf{313}, 276 (1993)

\bibitem{E852:2001ikk}
  E.~I.~Ivanov \textit{et al.} (E852 Collaboration),
  Observation of exotic meson production in the reaction $\pi^-p\to\eta^\prime\pi^-p$ at 18 GeV$/c$,
  Phys. Rev. Lett. \textbf{86}, 3977 (2001).

\bibitem{Amelin:2005ry}
  D.~V.~Amelin \textit{et al.} (E852 Collaboration),
  Investigation of hybrid states in the VES experiment at the Institute for High Energy Physics (Protvino),
  Phys. Atom. Nucl. \textbf{68}, 359 (2005).

\bibitem{Zaitsev:2000rc}
  A.~Zaitsev \textit{et al.} (VES Collaboration),
  Study of exotic resonances in diffractive reactions,
  Nucl. Phys. A \textbf{675}, 155 (2000).

\bibitem{E852:1998mbq}
  G.~S.~Adams \textit{et al.} (E852 Collaboration),
  Observation of a new $J^{PC}=1^{-+}$ exotic state in the reaction $\pi^-p\to\pi^+\pi^-\pi^-p$ at 18 GeV$/c$,
  Phys. Rev. Lett. \textbf{81}, 5760 (1998).

\bibitem{COMPASS:2009xrl}
  M.~Alekseev \textit{et al.} (COMPASS Collaboration),
  Observation of a $J^{PC}=1^{-+}$ exotic resonance in diffractive dissociation of 190 GeV$/c$ $\pi^-$ into $\pi^-\pi^-\pi^+$,
  Phys. Rev. Lett. \textbf{104}, 241803 (2010).

\bibitem{COMPASS:2021ogp}
  M.~G.~Alexeev \textit{et al.} (COMPASS Collaboration),
  Exotic meson $\pi_1(1600)$ with $J^{PC} = 1^{-+}$ and its decay into $\rho(770)\pi$,
  Phys. Rev. D \textbf{105}, 012005 (2022).



\bibitem{Baker:2003jh}
  C.~A.~Baker \textit{et al.},
  Confirmation of $a_0(1450)$ and $\pi_1(1600)$ in $\bar{p}p\to\omega\pi^+\pi^-\pi^0$ at rest,
  Phys. Lett. B \textbf{563}, 140 (2003).

\bibitem{Meyer:2010ku}
  C.~A.~Meyer and Y.~Van Haarlem,
  The status of exotic-quantum-number mesons,
  Phys. Rev. C \textbf{82}, 025208 (2010).

\bibitem{Meyer:2015eta}
  C.~A.~Meyer and E.~S.~Swanson,
  Hybrid mesons,
  Prog. Part. Nucl. Phys. \textbf{82}, 21 (2015).

\bibitem{Ketzer:2019wmd}
  B.~Ketzer, B.~Grube and D.~Ryabchikov,
  Light-meson spectroscopy with COMPASS,
  Prog. Part. Nucl. Phys. \textbf{113}, 103755 (2020).

\bibitem{Chen:2022asf}
  H.~X.~Chen, W.~Chen, X.~Liu, Y.~R.~Liu and S.~L.~Zhu,
  An updated review of the new hadron states,
  Rept. Prog. Phys. \textbf{86}, 026201 (2023).

\bibitem{BESIII:2022iwi}
  M.~Ablikim \textit{et al.} (BESIII Collaboration),
  Partial wave analysis of $J/\psi\rightarrow\gamma\eta\eta'$,
  Phys. Rev. D \textbf{106}, 072012 (2022),
  erratum: Phys. Rev. D \textbf{107}, 079901 (2023).

\bibitem{BESIII:2022riz}
  M.~Ablikim \textit{et al.} (BESIII Collaboration),
  Observation of an isoscalar resonance with exotic $J^{PC}=1^{-+}$ quantum numbers in $J/\psi\rightarrow\gamma\eta\eta'$,
  Phys. Rev. Lett. \textbf{129}, 192002 (2022),
  erratum: Phys. Rev. Lett. \textbf{130}, 159901 (2023).

\bibitem{Dong:2022cuw}
  X.~K.~Dong, Y.~H.~Lin and B.~S.~Zou,
  Interpretation of the $\eta_{1}$ (1855) as a $K\bar{K}_{1}(1400)+c.c.$ molecule,
  Sci. China Phys. Mech. Astron. \textbf{65}, 261011 (2022)

\bibitem{Yang:2022rck}
  F.~Yang, H.~Q.~Zhu, and Y.~Huang,
  Analysis of the $\eta_1(1855)$ as a $K\bar{K}_1(1400)$ molecular state,
  Nucl. Phys. A \textbf{1030}, 122571 (2023).

\bibitem{Yan:2023vbh}
  M.~J.~Yan, J.~M.~Dias, A.~Guevara, F.~K.~Guo and B.~S.~Zou,
  On the $\eta_{1}(1855)$, $\pi_{1}(1400)$, and $\pi_{1}(1600)$ as dynamically generated states and their SU(3) partners,
  Universe \textbf{9}, 109 (2023).

\bibitem{Wan:2022xkx}
  B.~D.~Wan, S.~Q.~Zhang and C.~F.~Qiao,
  Possible structure of newly found exotic state $\eta_1(1855)$,
  Phys. Rev. D \textbf{106}, 074003 (2022).

\bibitem{Qiu:2022ktc}
  L.~Qiu and Q.~Zhao,
  Towards the establishment of the light $J^{P(C)}=1^{-(+)}$ hybrid nonet,
  Chin. Phys. C \textbf{46}, 051001 (2022).

\bibitem{Chen:2022qpd}
  H.~X.~Chen, N.~Su and S.~L.~Zhu,
  QCD axial anomaly enhances the $\eta\eta^\prime$ decay of the hybrid candidate $\eta_{1}(1855)$,
  Chin. Phys. Lett. \textbf{39}, 051201 (2022).

\bibitem{Shastry:2022mhk}
  V.~Shastry, C.~S.~Fischer and F.~Giacosa,
  The phenomenology of the exotic hybrid nonet with $\pi_1(1600)$ and $\eta_1(1855)$,
  Phys. Lett. B \textbf{834}, 137478 (2022).

\bibitem{Wang:2022sib}
  X.~Y.~Wang, F.~C.~Zeng and X.~Liu,
  Production of the $\eta_1(1855)$ through kaon induced reactions under the assumptions that it is a molecular or a hybrid state,
  Phys. Rev. D \textbf{106}, 036005 (2022).

\bibitem{Huang:2022tpq}
  Y.~Huang and H.~Q.~Zhu,
  Revealing the inner structure of the newly observed $\eta_1(1855)$ via photoproduction,
  arXiv:2209.02879.

\bibitem{Shastry:2023ths}
  V.~Shastry and F.~Giacosa,
  Radiative production and decays of the exotic $\eta_1^\prime(1855)$ and its siblings,
  Nucl. Phys. A \textbf{1037}, 122683 (2023).

\bibitem{Chen:2022isv}
  F.~Chen \textit{et al.},
  $1^{-+}$ hybrid in $J/\psi$ radiative decays from Lattice QCD,
  Phys. Rev. D \textbf{107}, 054511 (2023).

\bibitem{Yu:2022wtu}
  Y.~Yu \textit{et al.},
  Investigating $\eta^\prime_{1}(1855)$ exotic states in $J/\psi\to\eta^\prime_{1}(1855)\eta^{(\prime)}$ decays,
  Phys. Lett. B \textbf{842}, 137965 (2023).

\bibitem{Barkai:1984ca}
  D.~Barkai, K.~J.~M.~Moriarty and C.~Rebbi,
  Force between static quarks,
  Phys. Rev. D \textbf{30}, 1293 (1984)

\bibitem{Itoh:1985wk}
  S.~Itoh, Y.~Iwasaki and T.~Yoshie,
  Static quark anti-quark potential with renormalization-group-improved lattice action,
  Phys. Rev. D \textbf{33}, 1806 (1986)


\bibitem{UKQCD:1998zbe}
  C.~R.~Allton \textit{et al.} (UKQCD Collaboration),
  Light hadron spectroscopy with $O(a)$-improved dynamical fermions,
  Phys. Rev. D \textbf{60}, 034507 (1999).

\bibitem{Karbstein:2018mzo}
  F.~Karbstein, M.~Wagner and M.~Weber,
  Determination of $\Lambda_{\overline{\textrm{MS}}}^{(n_f=2)}$ and analytic parametrization of the static quark-antiquark potential,
  Phys. Rev. D \textbf{98}, 114506 (2018)

\bibitem{Capitani:2018rox}
  S.~Capitani, O.~Philipsen, C.~Reisinger, C.~Riehl and M.~Wagner,
  Precision computation of hybrid static potentials in SU(3) lattice gauge theory,
  Phys. Rev. D \textbf{99}, 034502 (2019).

\bibitem{Schlosser:2021wnr}
  C.~Schlosser and M.~Wagner,
  Hybrid static potentials in SU(3) lattice gauge theory at small quark-antiquark separations,
  Phys. Rev. D \textbf{105}, 054503 (2022).

\bibitem{Braaten:2014qka}
  E.~Braaten, C.~Langmack and D.~H.~Smith,
  Born-Oppenheimer approximation for the XYZ Mesons,
  Phys. Rev. D \textbf{90}, 014044 (2014)


\bibitem{Salpeter:1951sz}
  E.~E.~Salpeter and H.~A.~Bethe,
  A relativistic equation for bound state problems,
  Phys. Rev. \textbf{84}, 1232 (1951).

\bibitem{Salpeter:1952ib}
  E.~E.~Salpeter,
  Mass corrections to the fine structure of hydrogen-like atoms,
  Phys. Rev. \textbf{87}, 328 (1952).

\bibitem{Lucha:1998ix}
  W.~Lucha and F.~F.~Schoberl,
  Semirelativistic treatment of bound states,
  Int. J. Mod. Phys. A \textbf{14}, 2309 (1999).

\bibitem{Isgur:1984bm}
  N.~Isgur and J.~E.~Paton,
  Flux-tube model for hadrons in QCD,
  Phys. Rev. D \textbf{31}, 2910 (1985)

\bibitem{Close:2003ae}
  F.~E.~Close and J.~J.~Dudek,
  Hybrid meson production by electromagnetic and weak interactions in a flux-tube model,
  Phys. Rev. D \textbf{69}, 034010 (2004)

\bibitem{Eshraim:2020ucw}
  W.~I.~Eshraim, C.~S.~Fischer, F.~Giacosa and D.~Parganlija,
  Hybrid phenomenology in a chiral approach,
  Eur. Phys. J. Plus \textbf{135}, 945 (2020).


\bibitem{Swanson:2023zlm}
  E.~S.~Swanson,
  Light hybrid meson mixing and phenomenology,
  Phys. Rev. D \textbf{107}, 074028 (2023).

\bibitem{Farina:2020slb}
  C.~Farina, H.~Garcia Tecocoatzi, A.~Giachino, E.~Santopinto and E.~S.~Swanson,
  Heavy hybrid decays in a constituent gluon model,
  Phys. Rev. D \textbf{102}, 014023 (2020).

\bibitem{Tanimoto:1982eh}
  M.~Tanimoto,
  Decay patterns of $q\bar{q}g$ hybrid mesons,
  Phys. Lett. B \textbf{116}, 198 (1982).

\bibitem{Tanimoto:1982wy}
  M.~Tanimoto,
  Decay of an exotic $q\bar{q}g$ hybrid meson,
  Phys. Rev. D \textbf{27}, 2648 (1983).

\bibitem{LeYaouanc:1984gh}
  A.~Le Yaouanc, L.~Oliver, O.~Pene, J.~C.~Raynal and S.~Ono,
  $q\bar{q}g$ hybrid mesons in $\psi \to \gamma$ + hadrons,
  Z. Phys. C \textbf{28}, 309 (1985).

\bibitem{Iddir:1988jd}
  F.~Iddir, A.~Le Yaouanc, L.~Oliver, O.~Pene, J.~C.~Raynal and S.~Ono,
  $q\bar{q}g$ hybrid and $qq\bar{q}\bar{q}$ diquonium interpretation of gams 1-+ resonance,
  Phys. Lett. B \textbf{205}, 564 (1988).

\bibitem{Ishida:1991mx}
  S.~Ishida, H.~Sawazaki, M.~Oda and K.~Yamada,
  Decay properties of hybrid mesons with a massive constituent gluon and search for their candidates,
  Phys. Rev. D \textbf{47}, 179 (1993).

\bibitem{Kalashnikova:1993xb}
  Y.~S.~Kalashnikova,
  Exotic hybrids and their nonexotic counterparts,
  Z. Phys. C \textbf{62}, 323 (1994).

\bibitem{Swanson:1997wy}
  E.~S.~Swanson and A.~P.~Szczepaniak,
  Decays of hybrid mesons,
  Phys. Rev. D \textbf{56}, 5692 (1997).

\bibitem{Iddir:2000yb}
  F.~Iddir and A.~S.~Safir,
  The decay of the observed $J^{PC} = 1^{-+}(1400)$ and $J^{PC} = 1^{-+}(1600)$ hybrid candidates,
  Phys. Lett. B \textbf{507}, 183 (2001).

\bibitem{Ding:2006ya}
  G.~J.~Ding and M.~L.~Yan,
  A candidate for $1^{--}$ strangeonium hybrid,
  Phys. Lett. B \textbf{650}, 390 (2007).

\bibitem{Iddir:2007dq}
  F.~Iddir and L.~Semlala,
  Hybrid states from constituent glue model,
  Int. J. Mod. Phys. A \textbf{23}, 5229 (2008).

\bibitem{Benhamida:2019nfx}
  A.~Benhamida and L.~Semlala,
  Hybrid meson interpretation of the exotic resonance $\pi_{1}(1600)$,
  Adv. High Energy Phys. \textbf{2020}, 9105240 (2020).

\bibitem{Isgur:1985vy}
  N.~Isgur, R.~Kokoski and J.~Paton,
  Gluonic excitations of mesons: Why they are missing and where to find them
  Phys. Rev. Lett. \textbf{54}, 869 (1985).

\bibitem{Close:1994hc}
  F.~E.~Close and P.~R.~Page,
  The production and decay of hybrid mesons by flux tube breaking,
  Nucl. Phys. B \textbf{443}, 233 (1995).

\bibitem{Barnes:1995hc}
  T.~Barnes, F.~E.~Close and E.~S.~Swanson,
  Hybrid and conventional mesons in the flux tube model: Numerical studies and their phenomenological implications,
  Phys. Rev. D \textbf{52}, 5242(1995).

\bibitem{Page:1998gz}
  P.~R.~Page, E.~S.~Swanson and A.~P.~Szczepaniak,
  Hybrid meson decay phenomenology,
  Phys. Rev. D \textbf{59}, 034016 (1999).

\bibitem{DeViron:1984svx}
  F.~De Viron and J.~Govaerts,
  Some decay modes of $1^{-+}$ hybrid mesons,
  Phys. Rev. Lett. \textbf{53}, 2207 (1984).

\bibitem{Zhu:1998sv}
  S.~L.~Zhu,
  Masses and decay widths of heavy hybrid mesons,
  Phys. Rev. D \textbf{60}, 014008 (1999).

\bibitem{Zhang:2002id}
  A.~l.~Zhang and T.~G.~Steele,
  Decays of the $\hat\rho(1^{-+})$ Exotic Hybrid and $\eta-\eta^\prime$ Mixing,
  Phys. Rev. D \textbf{65}, 114013 (2002).

\bibitem{Chen:2010ic}
  H.~X.~Chen, Z.~X.~Cai, P.~Z.~Huang and S.~L.~Zhu,
  Decay properties of the $1^{-+}$ hybrid state,
  Phys. Rev. D \textbf{83}, 014006 (2011).

\bibitem{Huang:2010dc}
  P.~Z.~Huang, H.~X.~Chen and S.~L.~Zhu,
  Strong decay patterns of the $1^{-+}$ exotic hybrid mesons,
  Phys. Rev. D \textbf{83}, 014021 (2011).

\bibitem{Huang:2016upt}
  Z.~R.~Huang, H.~Y.~Jin, T.~G.~Steele and Z.~F.~Zhang,
  Revisiting the $b_1\pi$ and $\rho\pi$ decay modes of the $1^{-+}$ light hybrid state with light-cone QCD sum rules,
  Phys. Rev. D \textbf{94}, 054037 (2016).

\bibitem{McNeile:2002az}
  C.~McNeile \textit{et al.} (UKQCD Collaboration),
  Hybrid meson decay from the lattice,
  Phys. Rev. D \textbf{65}, 094505 (2002)

\bibitem{McNeile:2006bz}
  C.~McNeile \textit{et al.} (UKQCD Collaboration),
  Decay width of light quark hybrid meson from the lattice,
  Phys. Rev. D \textbf{73}, 074506 (2006).

\bibitem{Woss:2020ayi}
  A.~J.~Woss \textit{et al.} (Hadron Spectrum Collaboration),
  Decays of an exotic $1^{-+}$ hybrid meson resonance in QCD,
  Phys. Rev. D \textbf{103}, 054502 (2021).

\bibitem{Page:2003qn}
  P.~R.~Page,
  Selection rules for $J^{PC}$ exotic hybrid meson decay in large $N_c$,
  Phys. Rev. D \textbf{70}, 016004 (2004).

\bibitem{HadronSpectrum:2012gic}
  L.~Liu \textit{et al.} (Hadron Spectrum Collaboration),
  Excited and exotic charmonium spectroscopy from lattice QCD,
  J. High Energy Phys.  \textbf{07}, 126 (2012).

\bibitem{Hayne:1981zy}
  C.~Hayne and N.~Isgur,
  Beyond the wave function at the origin: Some momentum dependent effects in the nonrelativistic quark model,
  Phys. Rev. D \textbf{25}, 1944 (1982)

\bibitem{Close:1987aw}
  F.~E.~Close and H.~J.~Lipkin,
  New experimental evidence for four-quark exotics: The Serpukhov $\phi\pi$ resonance and the GAMS $\eta\pi$ enhancement,
  Phys. Lett. B \textbf{196}, 245 (1987)

\bibitem{Burns:2006wz}
  T.~Burns and F.~E.~Close,
  Hybrid meson properties in Lattice QCD and Flux Tube Models,
  Phys. Rev. D \textbf{74}, 034003 (2006).

\bibitem{Ambrosino:2009sc}
  F.~Ambrosino \textit{et al.} (KLOE Collaboration),
  A global fit to determine the pseudoscalar mixing angle and the gluonium content of the $\eta^\prime$ meson,
  J. High Energy Phys. \textbf{07}, 105 (2009).

\bibitem{Suzuki:1993yc}
  M.~Suzuki,
  Strange axial-vector mesons,
  Phys. Rev. D \textbf{47}, 1252 (1993).

\bibitem{Tayduganov:2011ui}
  A.~Tayduganov, E.~Kou and A.~Le Yaouanc,
  The strong decays of $K_1$ resonances,
  Phys. Rev. D \textbf{85}, 074011 (2012).

\bibitem{Bibrzycki:2021rwh}
  L.~Bibrzycki \textit{et al.} (JPAC Collaboration),
  $\pi^-p\to\eta^{(\prime)}\, \pi^- p$ in the double-Regge region,
  Eur. Phys. J. C \textbf{81}, 647 (2021),
  Erratum: Eur. Phys. J. C \textbf{81}, 915 (2021).

\bibitem{BESIII:2020nme}
  M.~Ablikim \textit{et al.} (BESIII Collaboration),
  Future physics programme of BESIII,
  Chin. Phys. C \textbf{44}, 040001 (2020).

\bibitem{Belle-II:2018jsg}
  E.~Kou \textit{et al.} (Belle-II Collaboration),
  The Belle II physics book,
  Prog. Theor. Exp. Phys. \textbf{2019}, 123C01 (2019), Erratum: Prog. Theor. Exp. Phys. \textbf{2020}, 029201 (2020)

\bibitem{Dudek:2012vr}
  J.~Dudek \textit{et al.},
  Physics opportunities with the 12 GeV upgrade at Jefferson Lab,
  Eur. Phys. J. A \textbf{48}, 187 (2012).

\bibitem{Burkert:2018nvj}
  V.~D.~Burkert,
  Jefferson Lab at 12 GeV: The Science Program,
  Ann. Rev. Nucl. Part. Sci. \textbf{68}, 405 (2018).

\bibitem{PANDA:2009yku}
  W. Erni \textit{et al.} ($\bar{P}ANDA$ Collaboration),
  Physics performance report for PANDA: Strong interaction studies with antiprotons,
  arXiv:0903.3905.


\end{thebibliography}
\end{document}